\documentstyle[11pt]{article}
 \def\NPB#1#2#3{Nucl. Phys.  {\bf B#1} (19#2) #3}
 \def\PLB#1#2#3{Phys. Lett.  {\bf B#1} (19#2) #3}
 
 \def\PRD#1#2#3{Phys. Rev.  {\bf D#1} (19#2) #3}
 \def\PRL#1#2#3{Phys. Rev. Lett. {\bf#1} (19#2) #3}

 \def\ARNP#1#2#3{Ann. Rev. Nucl. Part. Sci. {\bf#1} (19#2) #3}

 \def\beq{\begin{equation}}
 \def\eeq{\end{equation}}
 \newcommand{\newc}{\newcommand}
 \newc{\sm}{Standard Model }
 \newc{\smd}{Standard Model}
 \newc{\dac}{discrete anomaly cancellation }
 \newc{\mup}{``$\mu$'' problem }
 \newc{\eps}{\epsilon}
 \newc{\barr}{\begin{eqnarray}}
 \newc{\earr}{\end{eqnarray}}
 \def\gappeq{\mathrel{\rlap {\raise.5ex\hbox{$>$}}
 {\lower.5ex\hbox{$\sim$}}}}
 \def\lappeq{\mathrel{\rlap{\raise.5ex\hbox{$<$}}
 {\lower.5ex\hbox{$\sim$}}}}
\def\l{\label}

\begin{document}

\begin{flushright}
 hep-ph/9510340 \\
 HD-THEP-95-31 \\
 IOA-95-322 \\
\end{flushright}

 \begin{center}
 {\bf NEUTRINO MASSES IN SUPERSYMMETRIC THEORIES} \\

 \vspace*{0.7 cm}
{\large {\bf George K. Leontaris$^{a}$ and Smaragda  Lola$^{a,b}$}}
 \end{center}
 \begin{center}
 \begin{tabular}{l}
 $^{a}$
 {\small
 Department of Physics, Ioannina University, Ioannina, Greece}\\
 $^{b}${\small Institut f\"{u}r
 Theoretische Physik, Univerisit\"at Heidelberg,Germany}\\
 \end{tabular}
 \end{center}

\vspace{0.3 cm}

\noindent
(Talk given by S. Lola at the 5th Hellenic School and Workshops
on Elementary Particle Physics, Corfu, September 1995).

\vspace{0.3 cm}

 \begin{center}
 {\bf ABSTRACT}
 \end{center}

\noindent
\small{
Extensions of the
Standard Model with additional $U(1)$ symmetries
can describe the hierarchy of fermion
masses and mixing angles, including neutrinos.
 The neutrino masses and
mixings are
determined up to a discrete ambiguity corresponding to the
representation content of the Higgs sector responsible for the
Majorana mass matrix. The solar and the atmospheric neutrino
deficits as well as the COBE data,
may be explained simultaneously in specific schemes
motivated by symmetries.
Using simple, analytic expressions, it is possible to
demonstrate the known effect that for small $\tan\beta$, phenomenologically
 interesting
neutrino masses would disturb the bottom-tau unification.
This however can be avoided in schemes with
a large $\mu-\tau$ mixing  in the charged leptonic sector.
On the other hand, for the large $\tan\beta$ regime, due to the fixed
point properties of the top as well as the bottom coupling
(which are described by analytic expressions,
for sufficiently large couplings) no modification to the bottom-tau
unification would occur. Still, large mixing in the
$\mu$-$\tau$ sector is desirable in this case as well,
in order to have a solution to the atmospheric neutrino
problem. In the same schemes, a relatively heavy
strange quark $\approx 200$ MeV is also predicted.}

 \section{Introduction}

Although the \sm describes successfully the strong and
electroweak  phenomena, there are still unanswered questions, related
to the origin of fermion masses and mixing angles.
An obvious possibility is that some symmetry,
 additional to that of the \sm
 is responsible for the
 pattern of masses and mixings
 that we see at low energies.
 And although unification on its own does not agree with
 experiment, when combined with supersymmetry
 it leads to very successful predictions
 for the  gauge couplings, the
 pattern and magnitude of spontaneous symmetry
 breaking at the elecroweak scale
 and the bottom -- tau ($ b$ --  $\tau$)  unification.
 A further indication that
 additional symmetries beyond the \sm
 exist, has been the observation that
 the fermion mixing angles and masses have values
 consistent with the appearance of
 ``texture'' zeros in the mass
 matrices \cite{textures,tex2}.

 On the other hand, neutrino data from various experiments
 seems to require certain
 mixings between various types of massive neutrinos.
 For these unknown neutrino masses and mixings,
 a similar hierarchy as the one for quarks and leptons
 may be expected to hold.
 The picture coming from the experiments is the following:
 The solar neutrino puzzle can be resolved through
 matter enhanced or vacuum oscillations with
 \begin{eqnarray}
 \sin2\theta_{ex} & = & (0.39-1.6)\times 10^{-2} ,
 \quad \Delta m^2= (0.6- 1.2)\times 10^{-5}
 eV^2 \nonumber \\
  \sin2\theta_{ex} & \ge & 0.75 ,
 \quad \Delta m^2  =  (0.5-1.1)\times 10^{-5} eV^2
 \label{eq:sol2}
 \end{eqnarray}
respectively \cite{solar}. On the other hand, the atmospheric
neutrino problem may be explained in
the case that large mixing and small
mass splitting involving the
muon neutrino exists \cite{atmo}.
Taking into account the
bounds from accelerator
and reactor disappearance
experiments one finds that
\begin{eqnarray}
sin^22\theta_{\mu \alpha}\geq 0.51-0.6
, \; \; \; \;
\delta m^2_{\nu_{\alpha}\nu_{\mu}}\leq
10^{-2} \; {\rm eV}^{2}
\end{eqnarray}
Finally,
if neutrinos are to provide
 a hot dark matter component, as COBE requires \cite{COBE},
 then the heavier neutrino(s) should have  mass in the
 range  $\sim (1-6)$ eV, the precise value depending on the number
of neutrinos
that have masses of this order of magnitude.

In what follows, we will discuss the expectations on these
masses and mixings, from textures predicted by $U(1)$ symmetries.

\section{Quark and Charged Lepton Masses}

We start by reviewing the construction of the model  of
quark and charged lepton masses, that has been
proposed by Ibanez and Ross \cite{IR}. The structure of the mass
matrices is determined by a family symmetry, $U(1)_{FD}$,
with the charge assignment of the \sm states given
in Table \ref{table:2a}. The need to preserve $SU(2)_L$
invariance requires left-handed up and down quarks (leptons)
to have the same charge. This, plus the additional
requirement of symmetric
matrices, indicates that all quarks (leptons) of the same i-th
generation transform with the same charge $\alpha _i(a_i)$.  The
full anomaly free Abelian group involves an additional family
independent component, $U(1)_{FI}$, and with this freedom
$U(1)_{FD}$ is made traceless without any loss of generality. Thus
$\alpha_3=-(\alpha_1+\alpha_2)$ and $a_3=-(a_1+a_2)$.

If the light Higgs, $H_{2}$, $H_{1}$, responsible for the up and
down quark masses respectively, have $U(1)$ charge so that only
the (3,3) renormalisable Yukawa coupling to $H_{2}$, $H_{1}$ is
allowed, then only the (3,3) element of the associated mass matrix
will be
non-zero. The remaining entries are generated when the
$U(1)$ symmetry is broken. This breaking is taken to be spontaneous
via \sm singlet fields,
$\theta,\; \bar{\theta}$, with $U(1)_{FD}$ charge -1, +1
respectively,
with equal vevs (vacuum expectation values). After this
breaking the mass matrix acquires its structure. For
example, the (3,2) entry in the up quark mass matrix appears at
$O(\eps^{\mid\alpha_2-\alpha_1 \mid} )$ because  U(1) charge
conservation allows only a coupling       $c^c t H_2(\theta
/M_2)^{\alpha_2-\alpha_1}, \;  \alpha_2>\alpha_1$ or   $c^ct
H_2(\bar{\theta} /M_2)^{\alpha_1-\alpha_2},\; \alpha_1>\alpha_2$.
Here
$\epsilon=(<\theta>/M_2)$ where $M_2$ is the unification mass
scale  which governs the higher dimension operators.
A different scale, $M_{1}$, is expected
for the down quark mass matrices (we come back to
this point below).
Suppressing unknown Yukawa couplings and their phases (which
are all expected to be of order unity),
one arrives at mass matrices of the
form
\begin{eqnarray}
\frac{M_u}{m_t}\approx \left(
\begin{array}{ccc}
\epsilon^{\mid 2+6a \mid } &
\epsilon^{\mid 3a \mid } &
\epsilon^{\mid 1+3a\mid }
\\
\epsilon^{\mid 3a \mid } &
\epsilon^{ 2 } &
\epsilon^{ 1 } \\
\epsilon^{\mid 1+3a \mid } &
\epsilon^{1 } & 1
\end{array}
\right)
, \; \; \; \;
\frac{M_d}{m_b}\approx \left (
\begin{array}{ccc}
\bar{\epsilon}^{\mid 2+6a \mid } &
\bar{\epsilon}^{\mid 3a \mid } &
\bar{\epsilon}^{\mid 1+3a \mid } \\
\bar{\epsilon}^{\mid 3a \mid } &
\bar{\epsilon}^{ 2 } &
\bar{\epsilon}^{ 1 } \\
\bar{\epsilon}^{\mid 1+3a \mid } &
\bar{\epsilon}^{1} & 1
\end{array}
\right)
\label{eq:massu}
\end{eqnarray}
where $\bar{\epsilon} = (\frac{<\theta >}{M_1})^{|\alpha_2-
\alpha_1|}$,
$\epsilon=(\frac{<\theta >}{M_2})^{|\alpha_2-\alpha_1|}$ and
$a=\alpha_1/(\alpha_2-\alpha_1)$.
In this simplest realisation, $h_b\approx h_{t}$
therefore we are in the large $\tan\beta$ regime of the
parameter space of the MSSM.
However, for a different $H_1$ and $H_2$ charge assignment,
or in the presence of a mixing with additional Higgs fields,
with the same $U(1)$
quantum numbers, it is possible to generate different $h_b$ and $h_{t}$
couplings, thus allowing for {\em any value of} $\tan \beta$.
With $a=1$,
$\sqrt{\eps}=\bar{\eps}=0.23$,
(implying that $M_2>M_1$),
 the matrices are in excellent agreement with the
measured values of the quark masses.
This relation for $\epsilon$ and
$\bar{\epsilon}$ will also be very
helpful  below, in order to determining the neutrino
mass spectrum.

The charged lepton mass matrix is determined in a similar way.
Requiring $m_b=m_{\tau}$ at unification, sets
$\alpha_1=a_1$ and then the charged lepton mass matrix is
\begin{equation}
\frac{M_l}{m_{\tau}}\approx \left (
\begin{array}{ccc}
\bar{\epsilon}^{\mid 2+6a-2b \mid } &
\bar{\epsilon}^{\mid 3a \mid } &
\bar{\epsilon}^{\mid 1+3a-b \mid } \\
\bar{\epsilon}^{\mid 3a \mid } &
\bar{\epsilon}^{ \mid 2(1-b) \mid } &
\bar{\epsilon}^{ \mid 1 -b \mid} \\
\bar{\epsilon}^{\mid 1+3a-b \mid } &
\bar{\epsilon}^{\mid 1-b \mid} &1
\end{array}
\right)
\label{eq:7}
\end{equation}
where $b=(\alpha_2-a_2)/(\alpha_2-\alpha_1)$.
{}From the two choices found in \cite{IR}
to lead to reasonable lepton masses, the one which
in principle leads to
the maximum mixing in the $\mu-\tau$ is the
choice $b=1/2$\footnote{
In \cite{IR}
a residual $Z_2$ discrete gauge symmetry after $U(1)$ breaking by
which the  electron
and muon fields get transformed by a factor $(-1)$, was imposed.
This resulted in entries
raised in a half-integer power being set to zero, eliminating
the (2,3) entries in the mass matrix at the GUT scale. However
this is not a necessary condition and once the
$Z_2$ symmetry is dropped,
the relevant (2,3) entries may be quite large.}.
We will come back to this point at a latter stage.

\section{First predictions for neutrino masses from symmetries}

The first step in an effort to describe neutrino
masses, is to determine the Dirac and heavy Majorana
mass matrices. Here, we look at what happens
in the case we add three generations
of right-handed neutrinos, which
will lead for predictions for light neutrino
masses through the See-Saw mechanism.
In such a scheme, the light Majorana neutrino
masses are given by
\begin{equation}
M^{eff}_{\nu}=M^D_{\nu}\cdot (M^M_{\nu_R})^{-1}\cdot M^D_{\nu}
\label{eq:meff}
\end{equation}
$SU(2)_L$ fixes the
$U(1)_{FD}$ charge of the left-handed neutrino states to be the
same as the charged leptons. The left- right- symmetry then fixes
the charges of the right-handed neutrinos as given in Table
\ref{table:2a} and therefore the neutrino Dirac mass is
\begin{equation}
\frac{M^D_{\nu_R}}{m_{\nu_{\tau}}}\approx \left (
\begin{array}{ccc}
{\epsilon}^{\mid 2+6a-2b \mid } &
{\epsilon}^{\mid 3a \mid } &
{\epsilon}^{\mid 1+3a-b \mid } \\
{\epsilon}^{\mid 3a \mid } &
{\epsilon}^{ \mid 2(1-b) \mid } &
{\epsilon}^{ \mid 1 -b \mid} \\
{\epsilon}^{\mid 1+3a-b \mid } &
{\epsilon}^{\mid 1-b \mid} & 1
\end{array}
\right)
\label{eq:nud}
\end{equation}
The scale of this mass matrix is the same as for the
up-quark mass matrices, similar to models
based on Grand Unified Theories.

Of course the mass matrix structure of neutrinos is more
complicated, due to the possibility of Majorana masses for the
right-handed components. These arise
from a term of the form
$\nu_R\nu_R\Sigma$  where $\Sigma$ is a $SU(3)\otimes
SU(2)\otimes U(1)$ invariant Higgs scalar field with $I_W=0$ and
$\nu_R$ is a right-handed neutrino.
The possible choices for the
$\Sigma$ $U(1)_{FD}$ charge
 will give a discrete spectrum of
possible forms for the Majorana mass,
$M^M_{\nu}$ \cite{DLLRS,LLR}.
For example if, in the absence of $U(1)_{FD}$ symmetry breaking,
the $\Sigma$ charge is the same as the $H_{1,2}$ doublet Higgs
charges, only the (3,3) element of $M_{\nu}$ will be non-zero.
Allowing for $U(1)_{FD}$ breaking by $<\theta>$ and
$<\bar{\theta}>$
the remaining
elements  in the Majorana mass matrix will be
generated in an analogous way to the generation of the Dirac mass
matrices\footnote{$<\Sigma>$ is significantly below the Planck
scale and thus $<\theta>$ dominates the $U(1)_{FD}$ breaking.}.

An important point is to determine the appropriate
expansion parameter and this brings us back to the
generation of the scales $M_1$ and $M_2$.
Consider a string compactification, which, besides
$H_1$ and $H_2$, leaves additional Higgs multiplets
$H_{1,2}^{a,b...},\bar{H}_{1,2}^{a,b...}$ light. The pairs of
Higgs fields in conjugate representations
are expected to acquire
gauge invariant masses, if there
is any stage of spontaneous symmetry breaking
at a scale $M$ below the
compactification scale, where $M=<\Phi>$
and $\Phi$ is a gauge invariant
combination of Higgs fields. However, there
may be further
sources of Higgs mass. The left-right
symmetry, essentially requires an extension of the gauge symmetry
to $SU(2)_L\otimes SU(2)_R$ at high scales. This will be broken
by a right-handed sneutrino vev in which case the mass degeneracy
of the $H_1$ and $H_2$ pair which transform as a $(1/2,1/2)$
representation under $SU(2)_L\otimes SU(2)_R$ can be split via
the coupling $<\tilde \nu_R>H_2H_x$ where $H_x$ transforms as
$(1/2,0)$. Such a contribution will generate $M_2\approx
<\tilde \nu_R>$, $M_1\approx M$.
Then, one expects that the
${\bar{\nu}}_R$ fields acquire a mass of $O(M_1)$ via a
$\Phi\nu{\bar\nu}$ coupling, implying
that the appropriate expansion parameter for the
Majorana  mass matrix is the same as that for the {\it down}
quarks and charged  leptons \cite{DLLRS}.

We may now compute the patterns of Majorana mass for the
different possible choices of $\Sigma$ charge.
These are given in Table 2 \cite{LLR}.
For $\alpha =1, \, \beta \equiv 1-b = 1/2$,
we can obtain the specific forms
for  Dirac and Majorana textures compatible with the correct
fermion mass predictions in the presence of the intermediate
neutrino scale.
In Table 3 we present the eigenvalues of
the heavy Majorana mass matrix for this choice
of $\alpha$ and $\beta$.
The eigenvalues of $m_{eff}$ are given
in Table 4.
The order of the
matrices in Tables 2 and 3 corresponds to
the one of Table 2.

The first point to note from these structures,
is that in none of the  cases does the light Majorana mass matrix
 have degenerate eigenvalues, which in the
past had been the most common assumption.
This occurs because the charges of the
right-handed neutrinos force the mass matrix entries to be of
different orders in powers of the expansion parameter
$\epsilon$. In the case where two components are coupled through an
off diagonal mass term as in cases 2, 4, 5 and 9,
two out of the three eigenvalues may be
approximately equal.
For the light Majorana neutrino masses, the structure of the
Dirac mass matrices results in an even larger spread.
{}From the values quoted in the introduction,
we see that in order to solve the
solar, atmospheric and dark matter problems simultaneously,
three nearly degenerate neutrinos of approximate
mass $1-2$ eV
are required \cite{psm}.  This
is not the case in the simplest scheme that we have been
discussing, without fine tuning. We will see below a more
complicated scheme, with
more than one $\Sigma$ fields, where this becomes possible.
Before doing so, however, let us consider the rest of the implications
that this simplest scheme has.

Besides the relative strength of the neutrino masses,
we would also like to know what are the
expectations for their absolute
magnitudes. This depends on the origin of the
field $\Sigma$.  If $\Sigma = \tilde{\bar{\nu}}_R
\tilde{\bar{\nu}}_R$  then the Majorana masses
are expressed in units
$<\tilde{\bar{\nu}}_R><\tilde{\bar{\nu}}_R>/M_c$ where $M_c$ is the
mass scale governing the appearance of higher dimension
operators, typically the string scale or $M_{Planck}$.
For a unification scale
$O(10^{16}GeV)$ it is reasonable to choose
$<\tilde{\nu}_R>=O(10^{16}GeV)$ leading to a scale $10^{13}-
10^{14}GeV$ for the Majorana mass scale. Then the
mass unit for the  light neutrinos is roughly
$O(4-0.4)eV$ for a top quark of $O(200)GeV$ \cite{DLLRS},
which is an interesting feature.
An additional interesting point is that
the mixing in the (2,3) entries
is of the correct order of magnitude
for a possible solution to the atmospheric
neutrino problem\footnote{
The mixing in the (1,2) sector is negligible.}.
Indeed, the diagonalising matrix
is given by
\beq
V\approx
 \left(
\begin{array}{ccc}
\sqrt{1-\bar{\epsilon}^2} &
\bar{\epsilon}  \\
-\bar{\epsilon} & \sqrt{1-\bar{\epsilon}^2}
\end{array}
\right)
\eeq

Then, the textures of Table 4 indicate towards two
possible solutions:
In solution (A),
one may fit the COBE
results and solve the solar neutrino problem,
while in solution (B) \cite{LLR}, it is possible to
obtain a
simultaneous solution to the solar
and the atmospheric
neutrino problems.
Whether we obtain the solution (A) or (B)
depends on the predicted mass splitting between
the two heavy neutrinos in
each of the six choices for the heavy Majorana mass matrix.
For a $\nu_{\tau} \approx 5$ eV,
 we obtain a muon neutrino mass
$m_{\nu_{\mu}} = m_{\nu_{\tau}} x_{i} = 0.06, 0.014$ and $0.003$
eV respectively, where
$x_{i} \equiv  e^6$, $e^8$
and $e^{10}$, is the splitting between the two larger
eigenvalues of $m_{eff}$.
This indicates that,
the matrices with a total splitting $e^{10}$ naturally
lead to a solution of the COBE measurements and
the solar neutrino problem.
On the other hand, for
$m_{\nu_{\tau}} \approx 0.1$ eV and  $x_{1} = e^6$,
$m_{\nu_{\mu}} = m_{\nu_{\tau}} x_{1} = 0.0012$ eV, which
may be marginally consistent with a solution to the
atmospheric and solar neutrino problems (remember
that coefficients of order unity have not
yet been defined in the solutions. This was recently done
in \cite{graham}, using infrared-fixed point arguments).
Since there are alternative schemes which lead to
an explanation of the COBE measurements, other
than hot and cold dark
matter\footnote{ For example, we have found that
domain walls may give structure at medium and
large scales if, either they are unstable, or the
minima of the potentials of the relevant scalar
field appear with different probabilities \cite{walls}.},
we believe that the scheme (B) should be
considered on equivalent grounds with the scheme (A),
which has been
discussed extensively in
\cite{DLLRS}.

\section{Solutions with three degenerate neutrinos}

In the previous section, the simplest scheme with a
$U(1)$ symmetry  has been considered and while two classes
of solutions were found, it has not been possible to solve
all three neutrino problems at the same time.
This problem is expected to disappear, once we
go to schemes with more than one $\Sigma$ fields.
However in this case the possible choices one can make
increase a lot. For this reason,
instead of searching a priori for a more sophisticated model that may
accommodate the experimental data, we will follow the opposite
procedure \cite{VV}:
We first consider
models that potentially allow the consistent incorporation
of all experimental data and look at the form that
the heavy Majorana mass matrix should have, and then see how
this mass matrix arises from symmetries.
To do so, we initially assume  a strong mixing in the
2-3 entries of the effective mass matrix.
This will then enable a solution of the atmospheric neutrino problem.
To simplify the analysis, we take the 1-2 and 1-3 mixing angles to be zero
in this simple example, assuming that
the MSW oscillations are generated by the
charged current interactions, as in \cite{DLT}.
Furthermore we take three nearly degenerate masses.
\noindent
{}From $M_{\nu_R} = m_D^{\dagger}\cdot m_{eff}^{-1}\cdot m_D$
and with the
mixing matrix
\beq
V_{\nu} = \left
(\begin{array}{ccc}
1 & 0 & 0 \\
0 & c_{1} & -s_{1} \\
0 & s_{1} & c_{1}
\end{array}
\right)\
\eeq
$m^{-1}_{eff} = V_{\nu}m_{eff}^{-1diag}V_{\nu}^T$ will have the form
\beq
m_{eff}^{-1} = \left (
\begin{array}{ccc}
\frac{1}{m_{1}} & 0 & 0 \\
0 & \frac{c^2_1}{m_2}+\frac{s_1^2}{m_3} &
c_1s_1(\frac{1}{m_2}-\frac{1}{m_3})
\\
0 &
c_1s_1(\frac{1}{m_2}-\frac{1}{m_3})
& \frac{c_1^2}{m_3}+\frac{s^2_1}{m_2}
\end{array}
\right)
\;
\equiv
\;
\left (
\begin{array}{ccc}
a & 0 & 0 \\
0 & b & d \\
0 & d & c
\end{array}
\right)\ .
\label{eq:form}
\eeq
where $m_{i}$ are the eigenvalues of
$m_{eff}$. Identifying the entries
gives
\begin{eqnarray}
\sin^{2}2\theta_{1} & = &
\frac{4 d^{2}}{(m_2^{-1}-m_3^{-1})^2}
\nonumber \\
m_1^{-1} & = & a
\nonumber \\
m_2^{-1} & = &
\frac{b}{2}
+ \frac{c}{2} + \frac{1}{2}
\sqrt{
b^{2}-2bc+c^2+4d^2}
\nonumber \\
m_3^{-1} & = &
\frac{b}{2}
+ \frac{c}{2} - \frac{1}{2}
\sqrt{
b^{2}-2bc+c^2+4d^2}\ ,
\end{eqnarray}
where $\theta_{1}$ is the
$\mu-\tau$ mixing angle.
The case of the absolute value of the three masses equal (i.e.
$m_1 = m_2$, $m_3 = -m_2$ is equivalent to
$b = c = 0, \; \; \; a = d$, therefore
$\sin^{2}2\theta_{1} = 1, \; \; \;
\theta = 45^{0}$.

Subsequently, we assume the very large class of models from
underlying unified models (such as strings and grand unified
theories, or partially unified models) which fix the
neutrino Dirac mass matrix to be proportional to the u-quark
mass matrix.
For example, the form of the heavy Majorana mass
matrix corresponding to an up-quark mass matrix
of the form \cite{giudi,RRR}
\beq
m_{\nu}^{D} =
\left (
\begin{array}{ccc}
0 & 0 & x \\
0 & x & 0 \\
x & 0 & 1
\end{array}
\right)\ ,
\eeq
is given by
\beq
\left (
\begin{array}{ccc}
x^{2}
(\frac{c_{1}^{2}}{m_{3}}
+ \frac{s_{1}^{2}}{m_{2}})
& x^{2}\frac{sin2\theta_{1}}{2}\left
(\frac{1}{m_2}-\frac{1}{m_3}\right)
& x
(\frac{c_{1}^{2}}{m_{3}}
+ \frac{s_{1}^{2}}{m_{2}}) \\
x^{2}\frac{sin2\theta_{1}}{2}\left
(\frac{1}{m_2}-\frac{1}{m_3}\right)
& x^{2}\left
(\frac{c_{1}^{2}}{m_{2}}
+ \frac{s_{1}^{2}}{m_{3}}
\right )
& x\frac{sin2\theta_{1}}{2}\left
(\frac{1}{m_2}-\frac{1}{m_3}\right)
\\
x^{2}\left
(\frac{c_{1}^{2}}{m_{3}}
+ \frac{s_{1}^{2}}{m_{2}}
\right )
& x\frac{sin2\theta_{1}}{2}\left
(\frac{1}{m_2}-\frac{1}{m_3}\right)
& \frac{x^2}{m_1} +
\frac{c_1^2}{m_3}+
\frac{s_1^{2}}{m_{2}}
\end{array}
\right)\ .
\eeq
For the above values of the three masses this becomes
\cite{VV}
\beq
M_{\nu_R}\ = \
\left (
\begin{array}{ccc}
0 & M_{N} x & 0 \\
M_{N} x & 0 & M_{N}  \\
0 & M_{N}   &  M_{N} x
\end{array}
\right)\ ,
\label{eq:pattern}
\eeq
where $M_{N} = xd \approx
10^{11}-10^{13}$
GeV. So we see that in this example the degeneracy of all three masses
and one large
mixing angle is consistent and may be understood in terms of texture zeroes
of the heavy Majorana neutrino mass matrix $M_{\nu_R}$.

A systematic study of such solutions has been carried in
\cite{VV}, where all possible cases
with at least one large mixing angle are given.
The quoted mass matrices may arise due to symmetries,
when more than one $\Sigma$ and $\theta$ fields are present.
To see how this occurs, let us note the following:
Assume the existence of a $\Sigma$ field with
a charge $-1$, which
makes the (2,3) entry unity. This leads to the relevant
heavy Majorana mass
matrices that we have already derived.
Suppose now that a second
$\Sigma$ exists, with quantum number $+2$.
This means that from the original matrix, the dominant element
will be the one with the biggest absolute power in $\bar{\epsilon}$
ie, the elements (2,2), (2,3) and (3,3) still would couple to
$\Sigma_{1}$ with charge $-1$, while the (1,2)  and (1,3)
will couple to $\Sigma_2$. Then the matrix will be
\begin{eqnarray}
M_u\approx \left(
\begin{array}{ccc}
0 & \epsilon^{-3+2} & \epsilon^{-4+2} \\
\epsilon^{-3+2} & \epsilon & 1 \\
\epsilon^{-4+2} & 1 & \epsilon^{-1}
\end{array}
\right)
\label{eq:mu}
\end{eqnarray}
This structure is similar to
 that of the example we just gave, with
the difference that the (2,2) element is of order $\epsilon$.
However this does not affect the predictions \cite{VV}, since it results to a
small
deviation from the picture that we have discussed.

 \section{RGE with RH-neutrinos: an analytic approach}

{}From the above it is clear that the interpretation of many
 important experimental facts is based on the existence of
 the right -- handed partners $\nu_{R_i}$ of the three left -- handed
 neutrinos, where the scale of mass of these particles is at least
three orders of
magnitude smaller
 than the gauge unification scale, $M_U$.
  Thus the running from the unification scale, $M_U\sim
 10^{16}$ GeV, down to the scale of $M_{\nu_R}$, must include
 radiative corrections  from  $\nu_R$ neutrinos. After that
 scale, $\nu_R$'s decouple from the spectrum, and an effective
 see -- saw mechanism is operative, c.f. eq( \ref{eq:meff}).
 In the presence of the right handed neutrino, the renormalization
 group  equations for the Yukawa couplings at the one--loop level,
 for the small $\tan\beta$ regime, where only the top and Dirac --
 type neutrino Yukawa couplings are large at the GUT scale,
 can be written in a diagonal basis as follows \cite{BP}
 \barr
 16\pi^2 \frac{d}{dt} h_t&=
 & \left(
 6 h_t^2  + h_N^2
   - G_U\right)  h_t, \label{eq:rg1}
 \\
 16\pi^2 \frac{d}{dt} h_N&=& \left(
  4h_N^2  + 3 h_t^2
   - G_N \right) h_N,
 \label{eq:rg2}  \\
 16\pi^2 \frac{d}{dt} h_b&=
 & \left(h_t^2 - G_D \right) h_b, \label{eq:rg3}
 \\
 16\pi^2 \frac{d}{dt} h_{\tau}&=&\left( h_N^2
  - G_E \right) h_{\tau},
 \label{eq:rg4}
 \earr
 Here, $h_\alpha$, $\alpha=U,D,E,N$, represent the
 $3 \otimes 3$ Yukawa matrices for the up and down quarks, charged
lepton and Dirac neutrinos, while $I$ is the $3 \otimes 3$ identity
matrix. Finally, $G_{\alpha}= \sum_{i=1}^3c_{\alpha}^ig_i(t)^2$ are
functions which depend on the  gauge couplings with the
coefficients $c_{\alpha}^i$'s given by \cite{DLT,VB}.
\barr
\{c_U^i \}_{i=1,2,3} &=& \left\{ \frac{13}{15},3,\frac{16}{3}
\right\}, \qquad \{c_D^i \}_{i=1,2,3} =
\left\{\frac{7}{15},3,\frac{16}{3} \right\}, \\  \{c_E^i
\}_{i=1,2,3} &=& \left\{ \frac{9}{5},3,0\right\}, \qquad \quad
\{c_N^i \}_{i=1,2,3} = \left\{ \frac{3}{5},3,0\right\}.
\earr
 Below $M_N$, the right handed neutrino decouples from the
 massless spectrum and we are left with the standard spectrum
 of the MSSM.  For scales $Q\le M_N$ the gauge and Yukawa
 couplings evolve according to the standard renormalisation
 group equations. To gain an insight into the effects of new couplings
 associated with  the
$\nu_R$ in the renormalisation group running we
integrate the above equations in the region
$M_N\le Q\le M_U$. We denote the top and $\nu_R$ Yukawas
at the  unification scale by
$h_G$, while the bottom and
 $\tau$ couplings are denoted with $h_{b_0},{h_{\tau_0}}$
respectively.
The top and neutrino
 Yukawa couplings at the
unification scale are equal, a relation which
 arises naturally not only in our case but in most of the Grand
 Unified Models which predict the existence of right handed
 neutrinos. Then \cite{LLR}
  \begin{eqnarray}
 h_t(t)&=&\gamma_U(t)h_G\xi_t^6\xi_N\\
 h_N(t)&=&\gamma_N(t)h_G\xi_t^3\xi_N^4\\
 h_b(t)&=&\gamma_D(t)h_{b_0}\xi_t\\
 h_{\tau}(t)&=&\gamma_E(t)h_{\tau_0}\xi_N
 \end{eqnarray}
 where the functions $\gamma_\alpha(t)$ and
 $\xi_{i}$ depend purely on
 gauge coupling constants and Yukawa couplings respectively,
 and are given by
 \barr
 \gamma_\alpha(t)&=&  \exp({\frac{1}{16\pi^2}\int_{t_0}^t
  G_\alpha(t) \,dt})= \prod_{j=1}^3 \left( \frac{\alpha_{j,0}}{\alpha_j}
 \right)^{c_\alpha^j/2b_j} \nonumber \\
 \xi_i&=& \exp({\frac{1}{16\pi^2}\int_{t_0}^t \lambda^2_{i}dt})
 \earr
 One then finds that
 \begin{equation}
 h_{b}(t_N)=\rho
 \xi_t\frac{\gamma_D}{\gamma_E}h_{\tau}(t_N) \label{rho}
 \end{equation}
 with $\rho=\frac{h_{b_0}}{h_{\tau_0}\xi_N}$.
 In the case of $b-\tau$ unification at $M_U$, we have
  $h_{\tau_0} =h_{b_0}$, while in the absence
 of the right -- handed neutrino $\xi_N \equiv 1$, thus
 $\rho =1 $ and the $m_b$ mass has the phenomenologically reasonable
prediction at low
 energies.
 In the presence of $\nu_R$ however, if $h_{\tau_0}
 =h_{b_0}$ at the GUT scale, the parameter $\rho$
 is no longer equal to unity since $\xi_N<1$. In fact the
 parameter $\xi_N$ becomes smaller for lower $M_N$ scales.
 Therefore, in order to restore the correct $m_b/m_{\tau}$ prediction at low
 energies we need $\rho =1$ corresponding to
 $  h_{b_0}=h_{\tau_0}\xi_N$.
 For $M_N\approx 10^{13}GeV$ for example and
 $h_G \ge 1$, we can estimate that
 $\xi(t_N)\approx 0.89$ thus, there is a corresponding
 $\sim 10\%$ deviation of the $\tau - b$ equality at the
 GUT scale \cite{LLR}, in agreement with the numerical results of
\cite{VB}.

In the case of  a large $\tan\beta$, a first thing to note is that
there are important corrections to the bottom mass
from one-loop graphs involving supersymmetric scalar masses and the $\mu$
parameter, which can be of the order of $(30-50)\%$
\cite{botcor}.
Moreover, even if one ignores these corrections, the effect
of the heavy neutrino scale is much smaller, since now
the bottom Yukawa coupling also runs to a fixed point, therefore
its initial value does not play an important role.
For example, for large $\tan\beta$, and $h_b\approx h_{t}$,
the product and ratio of the top and bottom
couplings, has been found in \cite{FLL} to be
\beq
h_t h_b \approx
\frac{{8\pi^2}\gamma_Q \gamma_D}{{7}\int \gamma_Q^2 d\,t},
\; \;  \;
\frac{h_t^2}{h_b^2}\approx \frac{\gamma_Q^2}{\gamma_D^2}
\label{fxd}
\eeq
indicating that one gets an approximate,
model independent  prediction for both couplings
at the low energy scale.
To see the effect of the neutrino scale to the
$b-\tau$ unification in this case,
we solved numerically the renormalisation group
equations. In the small $\tan\beta$ regime,
there exists a  parameter
space where the initial condition
$h_t = 2.0$ and $h_b= 0.0125$ lead to a factor $\xi_N = 0.86$,
for $M_N = 10^{12}$ GeV and an upper limit for the running bottom
mass $m_b=4.35$.
For the same parameter space,
when we set $h_b=2.0$, $\xi$ becomes $\xi_N=0.96$.
Moreover, again for the same
example, if we allow for a running bottom mass $m_b=4.4$, $\xi_N=0.99$.
For higher heavy neutrino scales,
the relevant effect is even smaller.

\section{Restoration of bottom -- tau unification}

Given the results of the previous section, it is natural to ask if
 Grand Unified models which predict the $b - \tau$
equality at the Unification scale,  exclude the experimentally
required and cosmologically interesting
region for the neutrino masses in the small $\tan\beta$ regime.
To answer this question, we should first recall that
the $b-\tau$ equality at the GUT scale refers to the
$(3,3)$ entries of the corresponding charged lepton and
down quark mass matrices. The detailed structure of the
mass matrices is not predicted, at least by the Grand Unified
Group itself, unless additional structure is imposed.
It is possible then to assume $(m_E^0)_{33}=
(m_D^0)_{33}$  and a specific structure of the corresponding mass
matrices such that after the diagonalisation at   the
GUT scale, the $(m^{\delta}_E)_{33}$ and $(m^{\delta}_D)_{33}$
entries are no-longer equal \cite{LLR}
\footnote{ An alternative solution occurs
in a class of models where the symmetries lead
to a neutrino Yukawa coupling
much smaller than the top one \cite{dp}
.}.

To illustrate this point, let us present
 here a simple $2\times 2 $ example \cite{LLR}.
 Assume a diagonal form of $m_D^0$ at the GUT scale ,
 $m_D^0 = diagonal (c m_0,m_0)$, while the corresponding
 entries of charged lepton mass matrix have the form
 \beq
 m_{E}^0 =
 \left (
 \begin{array}{cc}
 d & \tilde{\epsilon} \\
 \tilde{\epsilon} &  1
 \end{array}
 \right) m_0
 \eeq
 These forms of $m_D^0,m_E^0$ ensure that at the GUT scale
 $(m_D^0)_{33}= (m_E^0)_{33}$. However, at low energies
 one should diagonalize the renormalised Yukawa matrices
 to obtain the correct eigenmasses. Equivalently, one can
 diagonalise the quark and charged lepton Yukawa matrices
 at the GUT scale and evolve separately the eigenstates and
 the mixing angles. Since $m_D^0$ has been chosen diagonal,
 the mass eigenstates
 which are to be identified with the $s,b$ -- quark masses at
 low energies are given by
$ m_s=c \gamma_D m_0$ and
$ m_{b} =  \gamma_D  m_0 \xi_t$,
 with $m_0 = h_{b_0} \frac{\upsilon}{\sqrt{2}}cos\beta$.
 To find the charged lepton mass eigenstates we need first to
 diagonalise $m_E^0$ at $M_{GUT}$. We can obtain the following
 relations between the entries $\tilde{\epsilon} ,d$ of $m_E^0$ and the mass
 eigenstates $m_{\mu}^0,m_{\tau}^0$ at the GUT scale
 \barr
 d=(\frac{m_{\tau}^0 - m_{\mu}^0}{m_0}-1), \; \; \;
 \tilde{\epsilon}^2
  = (\frac{m_{\mu}^0}{m_0}+1) (\frac{m_{\tau}^0}{m_0}-1)
 \earr
 In the presence of right handed neutrinos, the evolution of
 the above $\tau -$ eigenstate down to low energies is
 described  by (\ref{eq:rg4}) with
 $m_{\tau_0}=h_{\tau_0}
 \frac{\upsilon}{\sqrt{2}}cos\beta$.  By simple comparison of
 the obtained formulae, we conclude that, to obtain the
 correct $m_{\tau}/m_b$ ratio at $m_W$ while preserving the
 $b - \tau$  unification at $m_{GUT}$, the $m_E^0$ entries
 should satisfy the following relations
 \barr
 \tilde{\epsilon} = \sqrt{\frac{1}{\xi_N}-1} ,&
 d \approx (\frac{1}{\xi_N}-1) = \tilde{\epsilon}^2
\l{eq:de}
 \earr
 The above result deserves some discussion.
 Firstly we see that it is possible to preserve $b - \tau$
 unification by assuming $2-3$ generation mixing in the lepton
 sector, even  if the effects of the $\nu_R$ states are included.
 Secondly, this mixing is related to a very simple parameter
 which depends only on the scale $M_N$ and the initial
 $h_N$ condition.
 The range of the coefficient $c$ in the diagonal form of the
$m_D^0$ -- matrix, can also be estimated
using the experimental values of the quark masses $m_s,m_b$.
An interesting observation is that the usual $GUT$ -- relation
for the $(2,2)$ -- matrix elements of the charged lepton and down quark
mass matrices, i.e., $(m_E)_{22}=-3 (m_D)_{22}$, which in our case
is satisfied for $c = -d/3$, implies here a relatively heavy strange
quark mass $m_s\sim 200$ MeV.
Smaller $m_s$ values are obtained if $-3c/d <1$
\cite{LLR}.

\section{Conclusions}

We have looked at the implications for
neutrino masses and mixings, coming from
 $U(1)$ symmetries,
in addition to the Standard Model gauge group.
We find that it is possible to explain the solar, the
atmospheric and the dark matter problems at the same time,
in schemes which can be derived from such symmetries.
Moreover, we have derived analytic expressions to describe
the fact that in the small $\tan\beta$ regime,
an intermediate neutrino scale would result to
deviations from the bottom-tau unification
(in the large $\tan\beta$ regime, one notices that
due to the top and bottom coupling fixed point properties,
no modification to the bottom-tau
unification would occur).
We proposed schemes where this deviation is avoided, by
considering
a large $\mu-\tau$ mixing  in the charged leptonic sector.
A relatively heavy
strange quark $\approx 200$ MeV is also predicted
in the framework of these models.

\pagebreak

\begin{table}
\centering
\begin{tabular}{|c |cccccccc|}\hline
   &$ Q_i$ & $u^c_i$ &$ d^c_i$ &$ L_i$ & $e^c_i$ & $\nu^c_i$ &
$H_2$ &
$ H_1$   \\
\hline
  $U(1)_{FD}$ & $\alpha _i$ & $\alpha _i$ & $\alpha _i$  & $a_i$
& $a_i$ & $a_i $ & $-2\alpha _1$ &  $-2\alpha _1$
\\
\hline
\end{tabular}
\caption{ $U(1)_{FD}$ charges. }
\label{table:2a}
%\end{table}

\vspace{0.3 cm}

%\begin{table}
\centering
\begin{tabular}{|c|c|} \hline
%%%%%%%%%%%%%1
$\left (
\begin{array}{ccc}
\bar\eps^{2\mid 3\alpha + \beta\mid } & \bar\eps^{3\mid\alpha\mid}
&\bar\eps^{\mid 3\alpha + \beta\mid } \\
\bar\eps^{\mid 3\alpha \mid } & \bar\eps^{2\mid \beta \mid} &
\bar\eps^{\mid \beta \mid}\\
\bar\eps^{\mid 3\alpha + \beta\mid } & \bar\eps^{\mid \beta \mid} & 1
\end{array}
\right)$
%%%%%%%%%%%%%2
 &
$\left (
\begin{array}{ccc}
\bar\eps^{3\mid 2\alpha + \beta\mid } & \bar\eps^{\mid 3\alpha + \beta\mid}
&\bar\eps^{\mid 3\alpha + 2\beta\mid } \\
\bar\eps^{\mid 3\alpha + \beta\mid } & \bar\eps^{\mid \beta \mid} & 1\\
\bar\eps^{\mid 3\alpha + 2\beta\mid }  & 1 &\bar\eps^{\mid \beta \mid}
\end{array}
\right)$
\\ \hline
%%%%%%%%%%%%%3
$\left (
\begin{array}{ccc}
\bar\eps^{2\mid 3\alpha + 2\beta\mid } &
 \bar\eps^{\mid 3\alpha + 2\beta\mid}
&\bar\eps^{3\mid \alpha + \beta\mid } \\
\bar\eps^{\mid 3\alpha + 2\beta\mid } &  1&\bar\eps^{\mid \beta \mid} \\
\bar\eps^{3\mid \alpha + \beta\mid }  &\bar\eps^{\mid \beta \mid}
&\bar\eps^{2\mid \beta \mid}
\end{array}
\right)$
%%%%%%%%%%%%%4
 &
$\left (
\begin{array}{ccc}
\bar\eps^{\mid 3\alpha + \beta\mid } &
 \bar\eps^{\mid \beta\mid} &1 \\
\bar\eps^{ \mid\beta\mid } & \bar\eps^{3\mid \alpha + \beta\mid }&
\bar\eps^{3\mid \alpha + 2\beta\mid } \\
 1 &\bar\eps^{3\mid \alpha + 2\beta\mid }
 &\bar\eps^{\mid 3\alpha + \beta \mid}
\end{array}
\right)$
 \\ \hline
%%%%%%%%%%%%%5
$\left (
\begin{array}{ccc}
1&\bar\eps^{\mid 3\alpha + 2\beta\mid } &
 \bar\eps^{\mid 3\alpha + \beta\mid}  \\
\bar\eps^{\mid 3\alpha +  2\beta\mid } &
 \bar\eps^{2\mid 3\alpha + 2\beta\mid }&
\bar\eps^{3\mid 2\alpha + \beta\mid } \\
\bar\eps^{\mid 3\alpha + \beta\mid} &
\bar\eps^{3\mid 2\alpha + \beta\mid }&
\bar\eps^{2\mid 3\alpha + \beta \mid}
\end{array}
\right)$
%%%%%%%%%%%%6
&
$\left(
\begin{array}{ccc}
\bar\eps^{\mid 3\alpha + 2\beta\mid }
& 1
 & \bar\eps^{\mid\beta\mid } \\
1  &
 \bar\eps^{\mid 3\alpha + 2\beta\mid }
 &\bar\eps^{\mid 3\alpha + \beta\mid } \\
\bar\eps^{\mid\beta\mid }  &
\bar\eps^{\mid 3\alpha + \beta\mid} &
\bar\eps^{\mid 3\alpha\mid}
\end{array}
\right)$
\\ \hline
\end{tabular}
\small{\caption{
General forms of heavy Majorana mass matrix textures.
The specific textures of the
text arise for $\alpha =1, \beta = 1/2$.}}
\label{table:maj}
%\end{table}

\vspace{0.5 cm}

%\begin{table}
\centering
\begin{tabular}{|c|c|} \hline
%%%%%%%%%%%%%1
$\left (
\begin{array}{ccc}
e^{10} &  &  \\
 & e^2 &  \\
 &  & 1
\end{array}
\right)$
%%%%%%%%%%%%%2
 &
$\left (
\begin{array}{ccc}
e^{15} &  &  \\
 & -1+e &  \\
 &  & 1+e
\end{array}
\right)$
\\ \hline
%%%%%%%%%%%%%3
$\left (
\begin{array}{ccc}
e^{16} &  &  \\
 & e^2 &  \\
 &  & 1
\end{array}
\right)$
%%%%%%%%%%%%%4
 &
$\left (
\begin{array}{ccc}
e^9 &  &  \\
 & -1-e^2 &  \\
 &  & 1+e^2
\end{array}
\right)$
 \\ \hline
%%%%%%%%%%%%%5
$\left (
\begin{array}{ccc}
e^{16} &  &  \\
 & e^{14} &  \\
 &  & 1
\end{array}
\right)$
%%%%%%%%%%%%6
&
$\left(
\begin{array}{ccc}
e^6 &  &  \\
 & -1-e^2 &  \\
 &  & 1+e^2
\end{array}
\right)$
\\ \hline
\end{tabular}
\caption{
Eigenvalues of Heavy Majorana mass matrix textures,
for $\alpha = 1$ and $\beta = 1/2$}
\label{table:majei}
%\end{table}

\vspace{0.5 cm}

%\begin{table}
\centering
\begin{tabular}{|c|c|} \hline
%%%%%%%%%%%%%1
$\left (
\begin{array}{ccc}
e^{26} &  &  \\
 & e^{10} &  \\
 &  & 1
\end{array}
\right)$
%%%%%%%%%%%%%2
 &
$\left (
\begin{array}{ccc}
e^{25} &  &  \\
 & e^9 &   \\
 &  & 1/e
\end{array}
\right)$
\\ \hline
%%%%%%%%%%%%%3
$\left (
\begin{array}{ccc}
e^{24} &  &  \\
 & e^8 &  \\
 &  & 1/e^2
\end{array}
\right)$
%%%%%%%%%%%%%4
 &
$\left (
\begin{array}{ccc}
e^{33} &  &  \\
 & e^{13} &  \\
 &  & 1/e^7
\end{array}
\right)$
 \\ \hline
%%%%%%%%%%%%%5
$\left (
\begin{array}{ccc}
e^{40} &  &  \\
 & 1/e^8 &  \\
 &  & 1/e^{14}
\end{array}
\right)$
%%%%%%%%%%%%6
&
$\left(
\begin{array}{ccc}
e^{32} &  &  \\
 & e^{6} &  \\
 &  & 1/e^6
\end{array}
\right)$
\\ \hline
\end{tabular}
\caption{
Eigenvalues of light Majorana mass matrix textures,
for $\alpha = 1$ and $\beta = 1/2$}
\label{table:majeil}
\end{table}

\end{document}